%% file: main.tex
\documentclass[10pt,conference,letterpaper]{IEEEtran}
\IEEEoverridecommandlockouts
\usepackage{cite}
\usepackage{amsmath,amssymb,amsfonts}
\usepackage{graphicx}
\usepackage{algorithmicx}
\usepackage{booktabs}
\usepackage[noend]{algpseudocode}
\usepackage{textcomp}
\usepackage{xcolor}
\usepackage{enumitem}
\usepackage{array}
\usepackage{footnote}
\usepackage{url}
\usepackage{balance}
\usepackage{multirow}
\usepackage{hyperref}
\usepackage{gensymb}

\makeatletter
\newcommand\fs@spaceruled{\def\@fs@cfont{\bfseries}\let\@fs@capt\floatc@ruled
  \def\@fs@pre{\vspace*{0.2cm}\hrule height.8pt depth0pt \kern2pt}%
  \def\@fs@post{\kern2pt\hrule\relax}%
  \def\@fs@mid{\kern2pt\hrule\kern2pt}%
  \let\@fs@iftopcapt\iftrue}
\makeatother

\begin{document}

\title{Technical Report:\\ Generating the WEB-IDS23 Dataset}

\author{ 
Eric Lanfer, Dominik Brockmann,  Nils Aschenbruck\\

Osnabrück University - Institute of Computer Science, Osnabrück, Germany\\
\{lanfer, dobrockmann, aschenbruck\}@uos.de\\
}
\IEEEoverridecommandlockouts


\maketitle

\begin{abstract}
Anomaly-based Network Intrusion Detection Systems (NIDS) require correctly labelled, representative and diverse datasets for an accurate evaluation and development. However, several widely used datasets do not include labels which are fine-grained enough and, together with small sample sizes, can lead to overfitting issues that also remain undetected when using test data. Additionally, the cybersecurity sector is evolving fast, and new attack mechanisms require the continuous creation of up-to-date datasets. To address these limitations, we developed a modular traffic generator that can simulate a wide variety of benign and malicious traffic. It incorporates multiple protocols, variability through randomization techniques and can produce attacks along corresponding benign traffic, as it occurs in real-world scenarios. Using the traffic generator, we create a dataset capturing over 12 million samples with 82 flow-level features and 21 fine-grained labels. Additionally, we include several web attack types which are often underrepresented in other datasets.
\end{abstract}

\input{introduction.tex}
\input{traffic_generation.tex}
\input{architecture.tex}
\input{Dataset.tex}
\input{caveats.tex}
\section{Conclusion}
In this technical report, we provide details on the creation of the WEB-IDS23 dataset. The dataset is publicly available and generated by a traffic generator developed by ourselves in a virtual environment. The aim of creating this new dataset is to encounter the lack of datasets with low numbers of attack samples to enable the community a balanced learning. Moreover, we include a set of web attacks that are used in real attack scenarios. We mixed the attack traffic with realistic benign traffic utilizing the same services as the attackers to form realistic scenarios.

The dataset is also pruned to some limitations, one big issue that it is synthetically generated and does not contain real-world traffic. Furthermore, only certain protocols are included, due to the high effort needed in implementing such actions into a traffic generator. Additionally, as described before in Sec.~\ref{sec:notes}, some attacks are only detectable by inspecting two flows, we did not implement a mechanism to trace back which flow was the result of a previous flow. 

In future work, when recording new datasets, it would be advisable to record the unencrypted payloads. They might be the only way to make some attacks differentiable from real traffic. This would also help to detect attacks that result in a second stream. By inspecting  the payloads, e.g., a revshell could be detected.

\section*{Acknowledgements}
We would like to thank Marty Schüller for his work on the traffic generator and for helping in recording and curating the dataset.
\balance
\bibliographystyle{IEEEtranS}
\bibliography{IEEEabrv, bibliography}

\end{document}

%% file: introduction.tex
\section{Introduction}
In order to develop anomaly-based network intrusion detection systems (NIDS), proper labelled, representative, and diverse datasets are essential. The recent years of IDS research brought up many datasets, which can be used to train and evaluate traffic classification models. But still, there is a lack of properly labelled data with recent attacks. The cybersecurity sector is changing very fast, new attack mechanisms evolve and due to the use of encryption, it gets harder to analyze traffic on the application layer. Therefore, it is important to create new datasets that can cover a diverse range of attacks and traffic characteristics, to enable researchers to develop and evaluate defense mechanisms, such as NIDS.

However, it is very difficult to obtain or create such datasets. When collecting real-world traffic, for example in a company or university network, high privacy standards make it almost impossible to collect datasets containing payload data. Besides this issue, the labelling problem of such datasets is even more impactful. The data needs to be manually labelled and inspected, since using a pattern-based intrusion detection system could miss malicious traffic or unknown attacks, where the system does not have any pattern for. Such labelling processes are very time-consuming and result in small sample numbers. For training machine learning models, however, large sample numbers are needed to cover a wide variety and avoid overfitting. Uneven represented classes and too small sample numbers are a problem observed in popular NIDS datasets \cite{Leung_Leckie,Portnoy01intrusiondetection}. Large datasets ensure model robustness by covering diverse attack scenarios and reducing the likelihood of overfitting to specific traffic patterns.

In several popular datasets widely used in intrusion detection research, overfitting is a big issue due to small sample sizes of certain attacks or the aggregation of such attacks into taxonomy super classes~\cite{zoghi2024unsw}. Without careful analysis and debugging of new models, such issues may not be visible, especially when solving the binary intrusion detection problem, where the samples need to be classified into benign and attack samples. For such debugging, accurate and fine-grained labels are needed, i.e., having only class labels indicating that traffic was malicious or not is not sufficient. Even taxonomy classes, such as \emph{DoS}, \emph{probe}, etc., are not fine-grained enough and may introduce further issues. This problem was already identified by John McHugh in his critiques of the DARPA dataset. He criticizes the grouping of certain attacks, e.g., in the \emph{DoS} category, summarizes types of attacks which are anatomically different \cite{mchugh2000testing}. The particular attack-types with an accurate description of how the attacks were performed are needed to enable domain experts to perform an insightful model performance analysis~\cite{10223401}.

Tackling the aforementioned issues, we developed a traffic generator, which is highly modular and configurable. We generate a dataset that includes a diverse range of attacks, fine-grained labels, and realistic traffic characteristics, addressing the key limitations of existing datasets. Additionally, the dataset includes several web attack types that are typically underrepresented in other datasets~\cite{ring2019survey}, along with corresponding benign actions targeting the same network services. This ensures that the dataset aligns with real-world traffic patterns, where benign and malicious actions coexist.

%% file: traffic_generation.tex
\section{Traffic Generation}
In order to have a wide variety of benign traffic and attack traffic, we developed a traffic generator based on Python. We decided to develop our own generator that allows us to orchestrate multiple attacks and also according benign traffic. Considering a company network, users will have benign interaction with the services provided on a network, these benign traffic usually mixes up with the traffic of attackers trying to attack such services. This overlap creates a realistic challenge for intrusion detection systems to detect the attacks inside the normal, benign noise. By implementing our own traffic generator, we can utilize the generator to execute benign actions that are closer to the attack patterns, e.g., utilizing the same web server endpoints. The generator supports multiple protocols, including HTTP(S), FTP, SMTP, SSH, ICMP, DNS, TCP, and UDP, ensuring broad coverage of typical network traffic scenarios. Additionally, the generator applies randomization techniques to introduce variability in request timings and traffic patterns, ensuring that generated traffic does not rely on static characteristics. This helps mitigate model overfitting to features like precise timing patterns that follow the specific generation process. Currently, the traffic generator is available on request to other researchers, it is planned to release it open source.

In the following, we will describe the two primary modes of the traffic generator: benign mode, where different actions are randomly triggered to simulate normal user behavior, and attack mode, where 13 different attack types can be executed.

\subsection{Benign Mode}
In benign mode, the traffic generator simulates normal user behavior across various protocols, including web browsing (HTTP(S)), file transfers via FTP, email exchanges through SMTP, and remote server interactions over SSH.

The traffic generator simulates benign HTTP(S) interactions using automated bots and a web crawler to replicate realistic web browsing behavior. Bots perform randomized actions on an OWASP Juice Shop instance, such as browsing pages, submitting feedback, and logging in or out. The web crawler complements this by navigating internal links from a list of URLs, scraping page content, and updating a graph-based crawl frontier to emulate natural user navigation. Both components introduce variability through randomized actions and delays. 
Similarly, benign FTP interactions involve randomized directory navigation, file uploads, and downloads, with variability in file names, sizes, permissions, and login attempts.
For SMTP, the traffic generator simulates randomized email exchanges by varying the subject, body length, and recipient count.
Finally, SSH interactions involve establishing remote sessions and executing random commands on servers, including delays and optionally simulated failures, creating realistic server access behaviors.

\subsection{Attack Mode}
In attack mode, the traffic generator simulates 13 different malicious activities across supported protocols. Table~\ref{tab:attacks} shows the different attacks grouped by the service.

For HTTP(S), various web-based attacks are executed on an OWASP Juice Shop instance, including SQL injection (targeting login and search functionalities, with optional payload obfuscation), cross-site scripting (XSS) through feedback forms, denial of service (DoS) via sqlmap, brute force login attempts using Hydra, server-side request forgery (SSRF) targeting predefined URLs, and reverse shell exploits leveraging server-side template injection (SSTI) to execute commands on the victim server. The implemented FTP attacks include fingerprinting via version detection and brute force attempts with randomized or injected credentials to simulate successful and unsuccessful login scenarios. For the attacks on the SMTP protocol, we focus on enumeration of users and fingerprinting of the mail server configuration. Furthermore, SSH attacks use brute force to compromise login credentials, incorporating successful breaches. Finally, miscellaneous attacks include host sweeps and port scans, with parameters such as target ranges randomized for variability. Each attack dynamically adjusts parameters, timing, and target specifics to ensure realistic simulation.

\begin{table}[]
\caption{Implemented attacks in the traffic generator}
\label{tab:attacks}
\resizebox{\columnwidth}{!}{%
\begin{tabular}{|l|l|l|l|}
\hline
\multicolumn{1}{|c|}{\textbf{Service}} & \multicolumn{1}{c|}{\textbf{Attack}} & \multicolumn{1}{c|}{\textbf{Alias}} & \multicolumn{1}{c|}{\textbf{Tool(s)}}                    \\ \hline
\textbf{FTP}                           &                                      &                                     &                                                          \\ \hline
                                       & Fingerprinting                       & ftp\_version                        & Metasploit (auxiliary/scanner/ftp/ftp\_version)   \\ \hline
                                       & Bruteforce                           & ftp\_login                          & Metasploit (auxiliary/scanner/ftp/ftp\_login)     \\ \hline
\textbf{HTTP/S}          &                                      &                                     &                                                          \\ \hline
                                       & Cross-site scripting         & xss                                 & Selenium WebDriver                                       \\ \hline
                                       & SQL-Injection                        & sqli                                & Selenium WebDriver, python-requests, sqlmap              \\ \hline
                                       & Denial of Service              & dos                                 & sqlmap                                                   \\ \hline
                                       & Bruteforce                           & bruteforce                          & Hydra                                                    \\ \hline
                                       & Server-side request forgery   & ssrf                                & Selenium WebDriver                                       \\ \hline
                                       & Reverse Shell                        & revshell                            & Selenium WebDriver, netcat                               \\ \hline
\textbf{SMTP}                          &                                      &                                     &                                                          \\ \hline
                                       & Fingerprinting                       & smtp\_version                       & Metasploit (auxiliary/scanner/smtp/smtp\_version) \\ \hline
                                       & User Enumeration                     & smtp\_enum                          & Metasploit (auxiliary/scanner/smtp/smtp\_enum)    \\ \hline
\textbf{SSH}                           &                                      &                                     &                                                          \\ \hline
                                       & Bruteforce                           & ssh\_login                          & Metasploit (auxiliary/scanner/ssh/ssh\_login)     \\ \hline
\textbf{Misc}  &                                      &                                     &                                                          \\ \hline
                                       & Portscan                             & portscan                            & Nmap (-sS flag)                                 \\ \hline
                                       & Hostsweep                            & hostsweep                           & Nmap (-sn and -Pn flags)                                          \\ \hline
\end{tabular}%
}
\end{table}

%% file: architecture.tex
\section{Architecture}
\begin{figure}
    \centering
    \includegraphics[width=\linewidth]{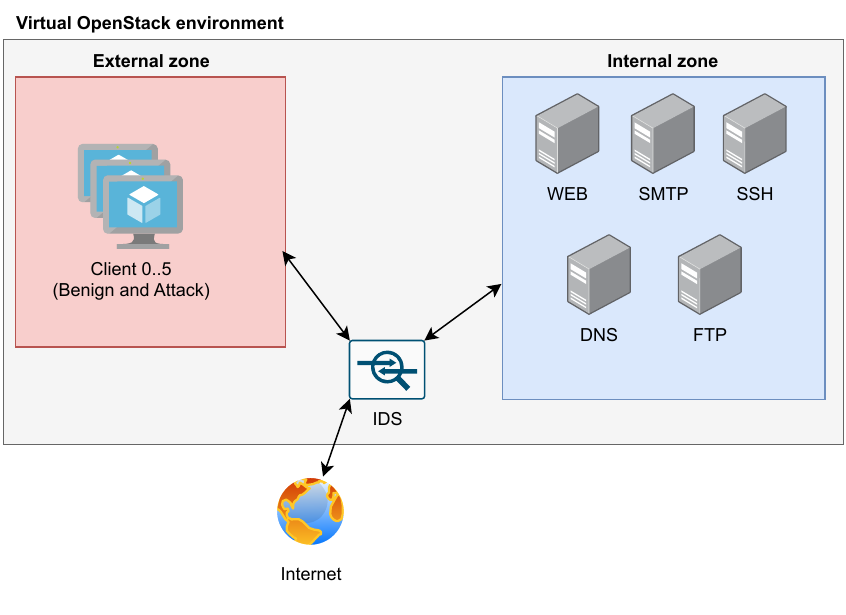}
    \caption{Virtual data capturing testbed}
    \label{fig:architecture}
\end{figure}

We utilize the traffic generator to generate a large dataset within a virtualized cloud environment based on OpenStack. Figure~\ref{fig:architecture} depicts the architecture of the testbed. It consists of two virtual networks interconnected via a router running Zeek\footnote{\url{https://zeek.org}} to record all traffic. The external zone contains five clients, while the internal zone hosts five servers. Both networks are connected to the internet to include realistic web traffic in the dataset. To prevent client fingerprinting and ensure data authenticity, each client operates in both normal and benign modes across different iterations.

%% file: dataset.tex
\section{Dataset}
We process the traffic recordings to create flow-level features. This results in a total of 82 features capturing various aspects, such as packet counts, inter-arrival times, payload characteristics as well as metadata about service type and duration. Each sample is labeled based on the logs generated by the traffic generator and result in 21 classes, including 20 attack classes and one benign class. In total, the dataset includes 12,059,749 samples. The dataset can be downloaded via the osnaData Repsoitory \cite{FK2/MOCIY8_2025}.

Table~\ref{tab:class_distribution} shows the distribution of the samples over the different classes. Among the attack classes, \emph{portscan} and \emph{hostsweep\_Pn} make up a large proportion of the dataset. In total, the dataset includes 825,187 benign samples that do not cover an attack. For each of the six web attacks (\emph{xss}, \emph{sql}, \emph{dos}, \emph{bruteforce}, \emph{ssrf}, \emph{revshell}), classes exist in both HTTP and HTTPS versions. For classification purposes, these classes might be kept as they are or merged, to apply classification independent of the protocol. For \emph{ssh\_login}, classes exist for successful as well as unsuccessful attempts, in the classes \emph{ssh\_login} and \emph{ssh\_login\_succesful}, respectively.

\begin{table}[tp]
\centering
\caption{Class Distribution}
\label{tab:class_distribution}
\begin{tabular}{|l|r|}
\hline
\textbf{Class}        & \textbf{Count}   \\ \hline
portscan                    & 5,046,406        \\ \hline
hostsweep\_Pn               & 3,492,290        \\ \hline
bruteforce\_http            & 912,503          \\ \hline
bruteforce\_https           & 865,126          \\ \hline
benign                      & 825,187          \\ \hline
ftp\_login                  & 468,275          \\ \hline
sql\_injection\_https       & 102,584          \\ \hline
dos\_http                   & 86,443           \\ \hline
sql\_injection\_http        & 74,300           \\ \hline
ssh\_login                  & 34,279           \\ \hline
ssh\_login\_successful      & 34,246           \\ \hline
dos\_https                  & 33,216           \\ \hline
hostsweep\_sn               & 22,637           \\ \hline
ftp\_version                & 11,688           \\ \hline
smtp\_version               & 11,353           \\ \hline
revshell\_https             & 9,404            \\ \hline
revshell\_http              & 8,549            \\ \hline
ssrf\_https                 & 6,656            \\ \hline
ssrf\_http                  & 5,509            \\ \hline
xss\_http                   & 4,558            \\ \hline
xss\_https                  & 4,533            \\ \hline
smtp\_enum                  & 7                \\ \hline
\end{tabular}%
\end{table}

We extract 80 flow-level features from the network traffic recordings using the \emph{Zeek FlowMeter} tool\footnote{\url{https://github.com/zeek-flowmeter/zeek-flowmeter}}, and additionally the service type and traffic direction for a total of 82 features. For most metrics, statistics such as the minimum, maximum, total, average, and standard deviation within a flow are computed. Features like \emph{flow\_duration}, the inter-arrival time of packets (e.g., \emph{fwd\_iat.min} and \emph{fwd\_iat.avg}) and the idle time capture temporal characteristics. Packet counts and directional statistics are given as the total packets in forward/backward direction, also in relation to the flow time and the ratio between forward and backward packets. The packet payloads are extracted as statistics about the payload length (e.g., \emph{fwd\_pkts\_payload.avg} and \emph{payload\_bytes\_per\_second}). Additionally, features about the header size, TCP control flags and bulk statistics are generated.
A more detailed description of each feature can be seen in the repository of the Zeek FlowMeter tool.

%% file: caveats.tex
\section{Notes on interpreting classifier results}
\label{sec:notes}
Some of the attacks presented in this dataset are not detectable by inspecting a single flow. Meaning, when a flow-based classifier detects such an attack correctly, it is likely that an overfitting issue occurred. The attacks \texttt{revshell} and \texttt{Server-side request forgery} are usually only successful when the victim server creates a new connection to a host specified in one of the attacks. This results in that proper detection is only guaranteed when at least two flows are analyzed. By only inspecting a single flow, the encrypted payload sent to a server should not be differentiable from a benign payload. Identifying these attacks requires either the inspection of unencrypted payloads or the inspection of resulting flows. However, during capturing the dataset we did not include a mechanism, that would be able to indicate, whether a stream resulted from a previous stream. With some uncertainty, it probably could be inferred, by inspecting the flows and matching on the flow that goes from the victim's IP to the attacker's IP, after a \texttt{revshell} samples is observed. For \texttt{ssrf} the target was randomly chosen out of a list of 5 hosts on the public internet, typically our victim server is not performing web requests. Therefore, this could be used as an indicator.